# Real-Time FPGA-Based Transformers & VLMs for Vision Tasks: SOTA Designs and Optimizations

Safa Mohammed Sali, Mahmoud Meribout, *Senior Member, IEEE* and Ashiyana Abdul Majeed

*Abstract*— Transformers and vision–language models (VLMs) have emerged as dominant architectures in computer vision and multimodal AI, offering state-of-the-art performance in tasks such as image classification, object detection, visual question answering, and caption generation. However, their high computational complexity, large memory footprints, and irregular data access patterns present significant challenges for deployment in latency- and power-constrained environments. Field-programmable gate arrays (FPGAs) provide an attractive hardware platform for such workloads due to their reconfigurability, fine-grained parallelism, and potential for energy-efficient acceleration. This paper presents a comprehensive review of design trade-offs, optimization strategies, and implementation challenges for FPGA-based inference of transformers and VLMs. We examine critical factors such as device-class selection, memory subsystem constraints, dataflow orchestration, quantization strategies, sparsity exploitation, and toolchain choices, alongside modality-specific issues unique to VLMs, including heterogeneous compute balancing and cross-attention memory management. Additionally, we discuss emerging trends in hardware–algorithm co-design, highlighting how innovations in attention mechanisms, compression techniques, and modular overlays can improve efficiency and adaptability. Practical considerations such as runtime flexibility, verification complexity, and the lack of standardized FPGA multimodal benchmarks are also addressed. Finally, future research directions are identified, focusing on scalable, portable, and reconfigurable FPGA solutions that can adapt to evolving model architectures while sustaining high utilization and predictable performance. Through this synthesis, the paper provides both a technical foundation and a forward-looking perspective, supporting researchers and engineers in bridging the gap between cutting-edge multimodal AI models and efficient, real-world FPGA deployment.

## I. INTRODUCTION

In recent years, transformer models have fundamentally reshaped the landscape of computer vision, particularly in tasks involving visual understanding and multimodal integration. Traditional computer vision techniques, which relied on carefully designed feature extractors, have given way to data-driven architectures that leverage attention mechanisms to capture complex relationships within and across data modalities. Vision Transformers (ViTs) exemplify this shift by applying self-attention to image patches, enabling models to grasp global context more effectively than conventional convolutional networks.

Building upon this, Vision-Language Models (VLMs) have emerged as powerful frameworks that fuse visual data with natural language, opening new avenues for tasks like image captioning, cross-modal retrieval, and visual reasoning. These models inherently demand substantial computational resources and memory bandwidth, posing significant challenges for deployment on devices with limited hardware capabilities.

The increasing computational and memory requirements of modern deep learning have fueled the development of specialized hardware accelerators, including GPUs, FPGAs, and ASICs. GPUs excel at parallel processing, making them the preferred choice for both training and inference tasks; however, their relatively high power consumption and broad-purpose design restrict their use in energy-constrained edge environments [1], [2]. ASICs deliver exceptional performance efficiency and compact silicon area, making them ideal for latency-sensitive and high-volume applications such as mobile and automotive systems, though their fixed hardware design limits flexibility [2], [3], [4]. Positioned between these extremes, FPGAs provide a versatile solution by combining reconfigurability with efficient pipelined processing, support for customized numerical precision, and adaptable dataflow architectures, which make them well-suited for real-time, low-latency inference in edge and streaming applications [5], [6], [7].

FPGAs offer a compelling platform for accelerating transformer and VLM workloads due to their adaptability, parallelism, and ability to implement customized dataflows. Recent advances in FPGA architectures, such as the integration of AI-specific compute units and enhanced memory hierarchies, support the efficient execution of transformer-based models. Additionally, model compression techniques—including pruning, quantization, and sparsity exploitation—play a vital role in tailoring transformers to meet the stringent latency and energy constraints typical of edge applications.

FPGAs have become a critical component in Advanced Driver Assistance Systems (ADAS) and edge vision applications. For instance, Subaru EyeSight employs Xilinx Zynq UltraScale+ MPSoCs to enable features such as adaptive

This work was supported by Khalifa University of Science and Technology. Safa Mohammed Sali is at Khalifa University of Science and Technology, Department of Computing and Information Engineering, Abu Dhabi, United Arab Emirates.

Mahmoud Meribout is at Khalifa University of Science and Technology, Department of Computing and Information Engineering, Abu Dhabi, United Arab Emirates (email: mahmoud.meribout@ku.ac.ae)

Ashiyana Abdul Majeed is at Khalifa University of Science and Technology, Department of Computing and Information Engineering, Abu Dhabi, United Arab Emirates (email: 100059454@ku.ac.ae).

cruise control, lane keeping, and collision avoidance [8]. Similarly, Continental's ARS540 system utilizes these FPGAs for 4D radar signal processing and robust multi-object tracking even in challenging environmental conditions [9]. In the lidar domain, companies like Ouster rely on Artix-7 and Zynq UltraScale+ MPSoCs to perform real-time obstacle detection and efficient point-cloud preprocessing [10], [11]. Beyond automotive, aerospace applications also benefit from FPGA technology; Northrop Grumman integrates Xilinx-based platforms such as BenERA and BenPRO to support demanding satellite processing tasks with high computational throughput [12].

This paper surveys the current state of FPGA-based acceleration strategies for transformer and vision-language models, focusing on architectural innovations, optimization techniques, and emerging software tools that facilitate rapid deployment. Key contributions include:

1. An in-depth examination of FPGA implementations targeting transformer and VLM architectures, highlighting trade-offs between computational efficiency and model fidelity.
2. Evaluation of hardware-aware optimizations tailored for attention mechanisms and multimodal fusion, emphasizing their impact on resource utilization and inference speed.
3. A review of cutting-edge development environments and frameworks that support the mapping of complex transformer and VLM models onto FPGA platforms, thereby accelerating research and application development.

Table 1 outlines the distinctions between our study and previous surveys, emphasizing its coverage of the latest (2025) FPGA-based Transformer and VLM research, the Versal device family, and sparsity optimization—areas that earlier transformer and LLM-focused reviews have not addressed. There are no literatures that addresses VLMs on FPGAs, making our paper stand out. Fig. 1 shows publication growth (2020-2025) in transformer and VLM based hardware techniques on FPGA, underscoring the increasing importance of this domain. By synthesizing developments in both machine learning and hardware design, this work aims to guide the creation of robust, low-power vision-language systems optimized for real-time edge processing. The literature review draws from a broad spectrum of recent publications, patents, and open-source projects, selected through keywords such as "2025," "real-time," "FPGA," "transformer/VLM detection," "transformer/VLM classification," and "transformer/VLM tracking".

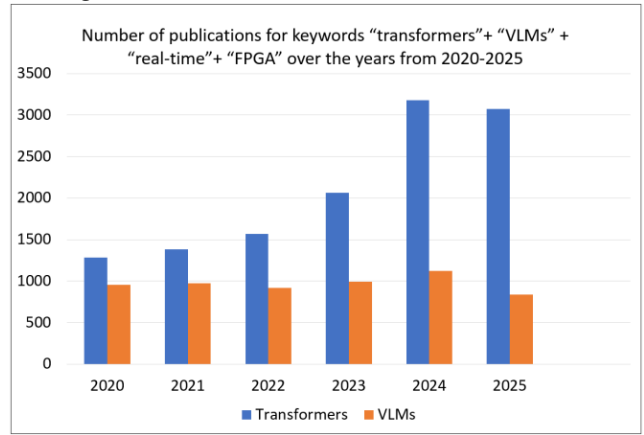

Fig. 1: Increasing trend in the field of transformer and VLM based literatures on FPGAs. The data is obtained from Semantic Scholar.

## II. FUNDAMENTALS AND BACKGROUND

### A. Computation Models of Transformer Models

Transformer models used for detection, classification and tracking share a consistent computation model centered around sequential tensor operations. Each layer in a transformer corresponds to specific stages of parallelizable computations, such as self-attention, matrix multiplication, and normalization. Architectures in this class process visual data as token sequences, enabling global context modeling through multi-head self-attention, but at the cost of substantial memory bandwidth and computational intensity. The processing pipeline begins with a CNN backbone (e.g., ResNet or DINOv2) extracting a high-dimensional feature map from the input image. This feature map is flattened into a sequence of tokens, each augmented with positional encodings to preserve spatial structure, and is then projected into queries (Q), keys (K), and values (V). Queries represent detection slots or object hypotheses, keys act as searchable indices containing spatial and semantic descriptors, and values store the associated feature content. The transformer encoder applies self-attention to these Q, K, and V sets, capturing long-range dependencies across the image. It consists of repeated blocks of multi-head self-attention (MHSA) followed by position-wise feedforward networks (FFNs), each wrapped with residual connections and layer normalization. The decoder extends this by introducing masked self-attention (to preserve causal order) and an additional encoder-decoder cross-attention mechanism. It performs cross-attention between the learned object queries and the encoder's key–value pairs, allowing each query to selectively retrieve information from the feature map before bounding box regression and classification. These sublayers are stacked multiple times depending on the model depth.

TABLE I
COMPARISON BETWEEN RECENT SURVEY LITERATURES

| Year | References | Vision Tasks | Model Types | FPGA Targets | Optimization Techniques | Sparsity | Highlights |
|---|---|---|---|---|---|---|---|
| 2025 | Ours | Detection, Classification, Tracking | Transformers, VLM | Xilinx Zynq, Intel, Versal, Kria, Versal | Quantization, Pruning, Pipelining, DSP Packing. | Yes | Focuses on object detection, classification, and tracking implementation of transformers and VLMs on FPGAs. Highlights sparsity and optimization. Covers hardware acceleration methods. |
| 2025 | Rati et al. [13] | - | Transformer | Xilinx, Intel | Pruning, Quantization | Yes | Focuses mainly on key architectural strategies, software challenges, and future research directions essential for efficient and scalable deployment. |
| 2025 | Kang etal. [14] | Detection | Transformer | Xilinx, Intel | Pruning, Quantization | Yes | Discussed FPGA-based hardware accelerators, optimization techniques, and compared FPGA architectures with ASIC |
| 2024 | Koilia et al. [15] | Detection and Classification | LLM | Xilinx Zynq, Alveo U200/U280, Virtex | Not covered | No | Discusses hardware acceleration of LLMs. |
| 2024 | Kachris etal. [16] | - | LLM | Xilinx, Intel | Not covered | No | Discusses hardware acceleration of LLMs. |

Table II summarizes the key computational operations, mathematical expressions, and complexity characteristics of each layer type commonly used in Transformer-based architectures. These layers are largely shared between encoder and decoder modules, though their ordering and masking differ. Among these, multi-head self-attention and feedforward layers are the most resource-intensive, with attention growing quadratically with input sequence length and feedforward layers scaling with the square of the hidden dimension. Attention layers also involve multiple matrix multiplications and softmax operations, making them bandwidth- and latency-sensitive. Meanwhile, feedforward layers contribute heavily to the total MAC operations due to their large dense weight matrices. In contrast, layers such as layer normalization and residual connections involve simple element-wise operations with minimal compute overhead.

All Transformer-based architectures rely on high-dimensional linear algebra operations, making them highly suitable for FPGA acceleration. The most computationally intensive components are the MHSA and FFN layers. Each attention head requires dense dot-product computations that can be efficiently parallelized. Similarly, FFN layers are composed of two large dense layers with intermediate nonlinear activations (e.g., GELU or ReLU) and typically dominate the total multiply–accumulate (MAC) operations.

Both MHSA and FFN layers also contribute substantially to memory bandwidth demands, owing to the large intermediate activations and parameter sizes. In contrast, layer normalization, applied after both attention and feedforward blocks, involves only element-wise mean and variance calculations, resulting in low computational complexity. Positional encoding, whether sinusoidal or learned, is added once at the embedding stage and does not significantly impact runtime complexity.

Unlike CNNs, Transformer models do not rely on spatial convolutions or pooling layers; instead, they model global and contextual dependencies through attention mechanisms. This allows them to scale flexibly across different input modalities (e.g., images, video, language) and resolutions. The modular and homogeneous nature of Transformer layers makes them particularly amenable to hardware-level parallelism and pipelining on reconfigurable platforms such as FPGAs.

Large Vision-Language Models (LVLMs), also called Multimodal LLMs (MLLMs) or Large Vision-Language Models (LVLMs), extend traditional LLMs to process both natural language and visual inputs, enabling unified reasoning across modalities (e.g., GPT-4V, LLaVA, MiniGPT) [17]. While both large language models and vision transformers (ViTs) share the transformer backbone, ViTs operate on visual tokens derived from image patches, whereas LLMs process linguistic tokens. In Multimodal LLMs, additional layers are introduced to accommodate and align multimodal inputs, leading to increased computational demands. While standard Transformer layers such as multi-head self-attention, feedforward networks, and normalization remain intact, multimodal LLMs extend the architecture with visual feature extraction blocks (e.g., CNNs or Vision Transformers), image patch projection layers, and cross-modal fusion mechanisms. These components are responsible for processing image tokens and aligning them with text embeddings either through early fusion (joint attention over concatenated tokens) or late fusion (modality-specific encoders with cross-attention). Such operations often increase the input sequence length and introduce extra attention heads, resulting in quadratic growth in attention complexity with respect to total token count.

Table III shows VLM layers apart additional to those adapted from the transformer model. Image features are typically projected using learned linear layers $W_{patch}$ and $b_{patch}$ to match the embedding size of textual tokens, which incurs additional matrix multiplications. Cross-attention layers, used to integrate information across modalities, replicate the computational cost of standard attention but between modality-specific representations, thus doubling certain matrix operations. Fusion modules may further introduce lightweight non-linearities or gating operations, which are relatively minor in compute but can impact latency when used repeatedly across layers.

While the core Transformer computational model is preserved, Multimodal LLMs exhibit higher compute and memory complexity due to increased token volume, repeated attention across modalities, and additional projection and fusion stages. These factors significantly affect throughput, memory bandwidth, and latency, making careful optimization essential for deployment on hardware accelerators. The patent [18] describes an LLM inference architecture that dramatically lowers peak memory requirements by employing layer-wise activation recomputation rather than storing all intermediate activations during a forward pass. At inference time, only minimal state is kept in memory, and needed activations are recomputed on-the-fly, trading modest extra computation for substantial memory savings, making it feasible to deploy larger language models on edge and mobile devices with tight RAM budgets.

TABLE II
COMPUTATIONAL MODELS OF TRANSFORMER LAYERS

| Layer Types (Functions) | Layer Name | Primary Operation | Computational Complexity | Remarks |
|---|---|---|---|---|
| Input Layers | Token Embedding | $x_i = E_{token}(t_i)$, where $t_i$ is token index, $E_{token}$ is token embedding matrix | $n \cdot d$ | Lookup or projection from vocabulary to dense space. Linear; requires memory for large vocab in NLP. |
| | Positional Encoding | $x_i = x_i + E_{pos}(t_i)$, where $E_{pos}$ is positional embedding for index $i$ | $n \cdot d$ | Adds positional info to token embeddings. Static (sinusoidal) or learned; no major compute overhead |
| Attention Layers | Linear Projections (QKV) | $Q = XW^Q, K = XW^K, V = XW^V$ | $3nd^2$ | Matrix multiplies to produce query, key, and value matrices. Highly parallelizable. |
| | Scaled Dot-Product Attention | $A = softmax(QK^T / \sqrt{d_k})\, V$ | $n^2 d$ | Dot products over sequence |
| | Causal attention (Masked Self-Attention) | $A = softmax\left(\frac{QK^T + M}{\sqrt{d_k}}\right) V$, where $M$ is Causal mask matrix $M_{ij} = -\infty$, for $j > i$ | $n^2 d$ | Prevents attending to future tokens. For decoder only |
| | Cross-Attention | $Q = XW^Q, K = ZW^K, V = ZW^V$, $A = softmax(QK^T / \sqrt{d_k})\, V$ | $n_{dec} \cdot n_{enc} \cdot d$ | Decoder attends to encoder |
| | Multi-Head Attention (MHSA) | $MHSA(X) = Concat(A_1, \ldots, A_h)W^O$, where $h$ is the number of attention heads, $W^O$ is output projection weight | $nd^2 + n^2 d$ | Combines outputs of multiple parallel attention heads; often the bottleneck |
| Feedforward Layers | Positionwise FFN | $FFN(X) = max(0, xW_1 + b_1)W_2 + b_2$, where $W_1$ and $W_2$ are weights for FFN linear layers, $b_1$ and $b_2$ are bias terms | $2n \cdot d \cdot d_{ff}$ | Two linear layers with activation. Commonly used in all transformer variants. |
| Normalization Layers | Layer Normalization | $\hat{x} = \frac{x - \mu}{\sqrt{\sigma^2 + \epsilon}} \cdot \gamma + \beta$, where, $x$ is input activation values to be normalized, $\mu$ and $\sigma^2$ are mean and variance respectively $\gamma, \beta$ learnable affine transform params, $\epsilon$ is small constant for numerical stability. | $n \cdot d$ | Applied after MHSA and FFN. Uses mean and variance across features (hidden dim). Lightweight element-wise ops. |
| Activation Layers | GELU | $GELU(x) = 0.5x \left(1 + tanh\left(\sqrt{2/\pi}\,(x + 0.0447x^3)\right)\right)$ | $n \cdot d$ | Nonlinear activation in FFN. Used in BERT and T5. Approximate sigmoid-weighted input. |
| | ReLU | $ReLU(x) = max(0, x)$ | $n \cdot d$ | Fast and common in earlier transformer variants. |
| | SiLU | $SiLU(x) = x \cdot \sigma(x)$, where $\sigma(x) = \frac{1}{1 + e^{-x}}$ | $n \cdot d$ | Smooth non-linearity. Used in Swin Transformer. |
| Skip/Utility Layers | Residual Connection | $y = x + F(x)$, where $x$ is the input tensor to the residual block and $F(x)$ denotes the transformation applied through a series of layers. | $n \cdot d$ | Element-wise addition. Used after attention and FFN sub-layers. No compute but adds to bandwidth. |
| | Dropout | $y_i = x_i \cdot Bernoulli(1 - p)$, where $p$ is dropout probability, i.e., fraction of units to drop (set to zero) during training | $n \cdot d$ | Applied during training only. Randomly zeroes elements; negligible in inference. |
| Output/Utility Layers | Classification / Projection | $y = softmax(XW^T + b)$, where $X \in \mathbb{R}^{n \times d}, W \in \mathbb{R}^{|V| \times d}$, $b \in \mathbb{R}^{|V|}, y \in \mathbb{R}^{n \times |V|}$, $y$ is output probability matrix, $b$ is the bias vector. | $n \cdot d \cdot |V|$ | Projects hidden states to output vocabulary or class space. Softmax applied across vocabulary dimension. Becomes expensive as $|V|$ increases. |

Where $X$ is input tensor (sequence of embedded tokens or features), $x_i$ is the individual token embedding at position $i$, $Q, K, V$ are Query, Key, Value matrices, $Z$ is encoder output tensor, $W^Q, W^K, W^V$ are projection matrices for $Q, K, V$, $A$ is attention score matrix after softmax, $K^T$ is transpose of the Key matrix $K$, $d_k$ is dimension of each attention head, $x$ is the token embedding passed through the two linear layers and activation function, $W^T$ is transposed weight matrix, $b$ is bias vector, $V$ is vocabulary size, d is model dimension, n is sequence length, $d_{ff}$ is the hidden size in FFN, $\mathbb{R}^{a \times b}$ denotes a matrix of real numbers with dimensions a×b.

TABLE III
COMPUTATIONAL MODELS OF MULTIMODAL LLMs

| Layer Types (Functions) | Layer Name | Primary Operation | Computational Complexity | Remarks |
|---|---|---|---|---|
| Input Layers | Image Patch Embedding | $V_i = Reshape(I)W_{patch} + b_{patch}$ | $n_v \cdot d$ | Converts $H \times W \times C$ image into $n_v$ patch tokens of dimension $d$. |
| Attention Layers | Vision Self-Attention | $A = softmax(QK^T / \sqrt{d_k}) V$, $Q = V_i W^Q, K = V_i W^K, V = V_i W^V$ | $n_v^2 d$ | MHSA over image tokens |
| | Text Self-Attention | $A = softmax(QK^T / \sqrt{d_k}) V$, $Q = XW^Q, K = XW^K, V = XW^V$ | $n_t^2 d$ | Standard Transformer attention for text tokens |
| | Cross-Attention (V→T) | $A = softmax(QK^T / \sqrt{d_k}) V$, $Q = XW^Q, K = V_i W^K, V = V_i W^V$ | $n_t \cdot n_v \cdot d$ | Text queries attend to visual tokens; core multimodal fusion step. |
| | Cross-Attention (T→V, optional) | $Q = V_i W^Q, K = XW^K, V = XW^V$, $A = softmax(QK^T / \sqrt{d_k}) V$ | $n_t \cdot n_v \cdot d$ | Optional bidirectional fusion; used in BLIP, Flamingo, etc. |
| Feedforward Layers | Feedforward Network (FFN) | $FFN(x) = W_2 \cdot \phi(W_1 x + b_1) + b_2$, where $W_1$ and $W_2$ are weights for FFN linear layers, $b_1$ and $b_2$ are bias terms | $n \cdot d^2$ | Two-layer Multi-Layer Perceptron (MLP) per token; activation $\phi$ = GELU or ReLU. |
| Normalization Layers | Layer Normalization | $LN(x) = \frac{x - \mu}{\sigma} \cdot \gamma + \beta$, where, $\mu$ and $\sigma$ are per token mean and standard deviation respectively. $\gamma, \beta$ learnable affine transform params. | $n \cdot d$ | Element-wise; used before or after attention/FFN layers. |
| Fusion / Utility Layers | Modality Gating / Fusion | $z = \alpha x + (1 - \alpha)y$, where $x$ is one input vector (e.g., original input, query, or hidden state), $y$ is another input vector (e.g., residual output, attention output), $\alpha$ is mixing coefficient (scalar or vector) between 0 and 1 (often learned or dynamically computed). | $n \cdot d$ | Combines text and vision embeddings with a learned weight α. |
| Output/Utility Layers | Classification / Projection | $y = softmax(XW^T + b)$, where $X \in \mathbb{R}^{n_t \times d}, W \in \mathbb{R}^{|V| \times d}$, $b \in \mathbb{R}^{|V|}, y \in \mathbb{R}^{n_t \times |V|}$, $y$ is output probability matrix, $b$ is the bias vector. | $n_t \cdot d \cdot |V|$ | Projects the final token embeddings to vocabulary or class logits. Dominated by matrix multiplication with large vocabulary size; often weight-tied with the input embedding matrix to reduce parameters. |

Where $X$ is input text tokens, $V_i$ is the input visual tokens, $W_{patch}$ is a learnable matrix that projects flattened image patches into the embedding dimension $d$, $b_{patch}$ is the patch projection bias vector (a learnable bias added to each projected patch embedding). $n_t$ is the number of text tokens, $n_v$ is number of visual tokens, and n is number of tokens processed in that layer. For text/self-attention, $n = n_t$ and for vision/self-attention, $n = n_t$. The remaining variables are same as in Table II.

*B. Datasets for detection, tracking and classification Models*

For transformer-based and vision–language models, curated datasets with rich, task-specific annotations are key to robust training, fair benchmarking, and meaningful comparison across detection, classification, and tracking tasks. The following are the datasets used for detection, classification and tracking in transformers and LLM models.

1) **Object Detection datasets:** Transformer detectors such as DETR [19] and DINO are typically benchmarked on MS COCO [20], KITTI [21], and nuScenes [22]. MS COCO remains a cornerstone dataset for detector development, offering over 118 000 training and 5 000 validation images across 80 classes, along with detailed instance segmentation masks and dense labelling. In the autonomous driving domain, benchmarks such as KITTI, with upwards of 7 000 stereo image pairs annotated, and BDD100K [23], featuring 70 000 driving video sequences and 100 000 detection-labeled images, introduce specialized real-world scenarios and challenges. nuScenes comprises 1,000 driving scenes with 3D bounding boxes and various sensor modalities. Large-scale image-text datasets like Open Images [24] (1.7 million images, 600 object classes), Visual Genome [25] (108,000 images with dense region captions) and MS COCO are used by LLM-inspired models like VLM, GLIP and Grounding DINO. These models are pre-trained on web-scale corpora like LAION-400M and LAION-5B [26], which contain 400 million to 5 billion image-text pairs, allowing for effective zero-shot object grounding and detection.

2) **Image Classification Datasets:** Vision transformers (ViT, Swin) and VLM classifiers often pretrain on ImageNet-21K (~14 M images, 21 k classes) or large image–text corpora (e.g., Conceptual Captions, CC12M), then fine-tune on ImageNet-1K (1.28 M/50 k train/val) and fine-grained sets like iNaturalist (2.7 M wildlife images, >10 k species) to achieve state-of-the-art accuracy with rich semantic alignment [27].

3) **Object Tracking Datasets:** On MOT17 and MOT20 [28], which concentrate on pedestrian multi-object tracking with thousands of frames and identity labels, transformer-based trackers like TransT and STARK have shown excellent performance. Large-scale datasets such as TAO [29] (2,907 videos, 833 object classes) and nuScenes Tracking [22] provide a variety of scenes and multi-object annotations for assessing tracking models inspired by transformers and LLMs. Large training corpora and pre-training on LAION-style datasets are advantageous for TrackFormer and MOTR, which combine detection and tracking in an end-to-end fashion. They then refine their performance on video-centric benchmarks such as KITTI Tracking [30], Argoverse [31], and TAO.

*C. Performance Metrics for Real-Time Object Detection*

A comprehensive set of evaluation criteria is essential for objectively comparing models designed for object detection, classification, and tracking. These include the following:

1) **Accuracy Measures:** For object detection, mean Average Precision (mAP) is the standard, summarizing precision–recall behavior over multiple IoU thresholds:

$$mAP = \frac{1}{N}\sum_{i=1}^{N} AP_i \qquad (1)$$

where $AP_i$ is the average precision for class i, and N is the number of object classes.

In multi-object tracking (MOT), performance is often captured by MOTA (Multiple Object Tracking Accuracy (MOTA) and MOTP (and Multiple Object Tracking Precision). MOTA accounts for missed detections, false alarms, and identity switches:

$$MOTA = 1 - \frac{\sum_t (FN_t + FP_t + IDSW_t)}{\sum_t GT_t} \qquad (2)$$

where $FN_t$, $FP_t$, and $IDSW_t$ denote the number of false negatives, false positives, and identity switches at time t, respectively, and $GT_t$ is the number of ground-truth objects

2) **Compute Complexity:** Model workload is often expressed in GFLOPs (or GMACs for reduced-precision arithmetic). For example, DETR framework using a ResNet-50 backbone requires on the order of 120 GFLOPs per inference on a 640 × 640 input image. It grows to about 180 GFLOPs at 800 × 800 resolution, since FLOPs scale roughly with the square of the image side length [32]. Efficient grounding DINO, a multi-modal LLM, requires 309.12 GFLOPs on DOTA dataset [33].

3) **Energy & Power Metrics:** Especially for edge applications, power efficiency (in GOPS/W) is a critical benchmark: Xilinx Zynq7 Z100 FPGAs achieve about 16.7 GOPS/W, compared with 6.32 GOPS/W for the NVIDIA RTX 3090 [34].

While peak tera-operations-per-second (TOPS) figures are often highlighted as a headline benchmark for AI-centric hardware, they rarely reflect actual deployed performance. Achieving this theoretical maximum requires full utilization of the tensor compute resources—an ideal that is seldom reached in practical workloads. Real-world efficiency is constrained primarily by how effectively the workload can be mapped onto the available compute fabric and by system-level factors such as the latency and bandwidth limits of moving data on and off the device.

## III. FPGA BASICS

From their origins as simple grids of configurable logic in the 1980s, FPGAs have grown into fully heterogeneous platforms that blend custom logic fabric with embedded processors and dedicated neural-processing blocks. Modern FPGA families generally fall into three classes: pure LUT-DSP based devices, SoC-based FPGAs; and SoC models augmented by fixed-function AI engines designed for high-throughput tensor operations. Both Intel (formerly Altera) and Xilinx (now under AMD) have championed these technologies for transformer- and VLM-based tasks such as object detection, classification, and tracking, leveraging increasingly sophisticated toolchains.

*A. Classical LUT and DSP blocks FPGAs*

Traditional LUT–DSP FPGAs consist of a two-dimensional array of Configurable Logic Blocks (CLBs) interconnected through a programmable routing network. Each CLB typically contains look-up tables (LUTs), flip-flops, and multiplexers, with the LUT functioning as a small truth table capable of implementing any Boolean expression involving up to $K$ inputs. The LUT contents are stored in configuration memory, allowing designers to realize bit-level custom datapaths and arbitrary quantization formats such as INT2, INT4, INT6, INT8, INT16, or even non-uniform encodings. This fine-grained control makes LUT-centric devices well-suited for exploring unconventional numerical formats, precision scaling, and quantization-aware design. However, in the absence of large dedicated arithmetic units, operations on wider bit-widths (e.g., INT16 or FP16) must be fully mapped into LUT fabric, which can be both area- and latency-intensive. To address compute-intensive tasks, most devices also embed DSP slices and block RAM (BRAM) to accelerate multiply–accumulate operations, often supporting operand sizes up to 54 × 54 bits [35]. Examples include the Lattice iCE40 series, Xilinx Spartan-7, and Intel Stratix 10 families, which prioritize architectural flexibility over peak throughput, making them effective for prototyping, algorithm exploration, and energy-conscious embedded designs.

LUT–DSP FPGAs are well-suited for time-critical workloads where deterministic response and reliable timing closure are essential. A representative device is the Artix UltraScale+ XA AU7P [35], a 9 × 9 mm automotive-qualified FPGA manufactured using 16 nm FinFET technology. It is tailored for dependable vision processing in demanding environmental conditions. The UltraScale+ architecture employs a column-and-grid organization, grouping CLBs, DSPs, block RAM (BRAM), I/O resources, and transceivers into uniform clock regions, as illustrated in Fig. 2. Low-skew interconnect ensures straightforward timing closure and supports microsecond-level reaction times.

The AU7P features 6-input LUT-based CLBs alongside 216 DSP48E2 slices. Each DSP slice implements a 27 × 18 multiplier with a 48-bit accumulator, and supports SIMD operation modes such as dual 24-bit or quad 12-bit, enabling efficient INT8/INT16 execution for CNN and DSP kernels.

Floating-point arithmetic (FP16/FP32) is available through Vivado Floating-Point IP cores, enabling mixed-precision implementations when higher numerical range is needed, at the cost of additional area and power usage. This balanced resource profile supports compact, power-conscious CNN inference in safety-critical automotive applications. Figure 3 presents the device's top-level resource arrangement.

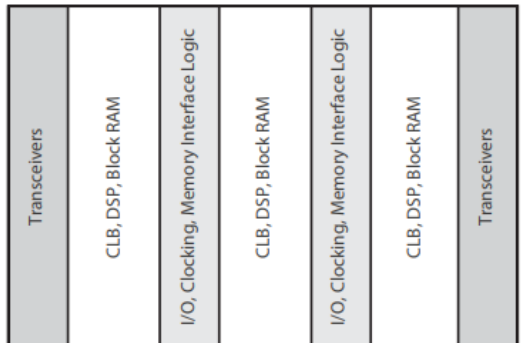

Fig. 2: FPGA with Columnar Resources from [35]

In these architectures, AI accelerators are typically composed of Processing Elements (PEs), which are small compute nodes that integrate local memory (registers or BRAM) with multiply–accumulate (MAC) hardware, delivering two FLOPs per MAC. When PEs are organized into a 2D systolic array, activations propagate vertically, weights flow horizontally, and partial sums accumulate along diagonals. This layout offers substantial spatial parallelism and on-chip data reuse [36]. Performance can be further improved through loop unrolling, which exposes parallelism by replicating loop bodies [37]; tiling, which divides large feature maps to fit within BRAM; and deep pipelining with an initiation interval (II) of one, ensuring a fresh input is processed every clock cycle. Double buffering of feature maps sustains throughput, while MAC operations map to DSP slices and weight/activation storage resides in BRAM, forming the computational backbone for CNN acceleration on LUT–DSP platforms [38], [39], [40].

*B. System on Chip (SoC) FPGAs*

Recent generations of Xilinx and Intel SoC FPGAs merge fixed ARM-based processing subsystems with the programmable logic (PL) fabric, enabling a heterogeneous computing platform where sequential or control-oriented workloads execute on CPUs while massively parallel tasks are accelerated in hardware. In Xilinx devices, this processing subsystem (PS) is branded as the Application Processing Unit (APU), whereas Intel refers to it as the Hard Processor System (HPS). Both implementations incorporate ARM Cortex-A series cores, I/O peripherals, memory interfaces, and DMA controllers, linked to the PL via multiple AXI channels—AXI-GP for low-latency control, AXI-HP for high-bandwidth transfers, and AXI-ACP for cache-coherent access. Representative examples include Intel's Agilex SoC [41], Xilinx Zynq UltraScale+ MPSoC [42], and the earlier Zynq-7000 family [43].

On the memory side, these SoCs typically support LPDDR3/4 or DDR4/5, with top-end configurations such as Agilex F-Series achieving up to 230.4 Gbps throughput using a 72-bit DDR4-3200 interface . Many variants integrate multimedia capabilities such as HDMI interfaces and hardware H.264 video decoding, making them well-suited for edge AI video analytics.

Within the programmable logic, high-performance DSP slices form the computational backbone for AI workloads. Xilinx's DSP48E1/E2 blocks, featuring 25×18 multipliers, pre-adders, 48-bit accumulators, and SIMD modes (dual 24-bit or quad 12-bit), enable efficient INT8/INT16 multiply-accumulate operations and can be extended to FP16/BF16. Intel's Agilex DSP blocks, in contrast, scale up to 54×54 multipliers and integrate with Adaptive Logic Modules (ALMs) and M20K embedded memories, supporting INT4 through FP32 arithmetic [44].

Although these DSP blocks are tuned for common data widths such as INT8 (dominant in machine learning inference), custom precisions like INT5 or INT7 can still be implemented using LUT resources, albeit with a penalty in latency and routing efficiency. Software tooling plays a pivotal role here. For instance, Xilinx's FINN [45] framework targets quantization-aware designs with bitwidths down to 1-bit, while the Vitis AI stack deploys INT8-optimized models to a soft Deep Learning Processing Unit, such as DPUCZDX8G [46].

The DPU itself is a LUT-DSP systolic array that connects to the APU through AXI interconnects, as depicted in Fig. 3. An instruction fetch stage feeds a high-performance scheduler, which maps CNN layers at compile-time onto a Hybrid Computing Array. Hybrid Computing Array is a grid of processing elements optimized for parallel MAC execution. Data is streamed from external DDR via a high-speed data path into a global on-chip buffer that stores intermediate activations, weights, and feature maps, ensuring continuous operation of the PEs. The architecture supports layer fusion (e.g., convolution + ReLU) to reduce intermediate memory transfers and improve efficiency. Among its variants, the B4096 configuration achieves 4,096 INT8 MACs per cycle, suitable for high-throughput CNN inference in real-time systems [47].

The DPU built fully on PL, enables substantial parallelism for CNN workloads but comes at the cost of higher logic resource usage and introduces latency of several tens of cycles per layer, depending on both network depth and memory bandwidth. Through integration with Vitis AI, it facilitates hardware-agnostic CNN deployment without requiring designers to work directly in HDL. An additional benefit of SoC-based FPGAs is their ability to execute real-time task scheduling within the Processing System (PS). Leveraging ARM cores, real-time operating systems such as FreeRTOS or embedded Linux enhanced with PREEMPT-RT patches can perform deterministic handling of time-critical tasks, manage interrupts, and control peripherals [48]. Nonetheless, existing literature has paid little attention to the use of RTOS in managing PS-PL coordination for AI-driven video applications, leaving an opportunity for research—particularly in deploying multiple AI models with varying priority levels and bounded worst-case scenario times (WSCT).

Intel's Agilex 3 FPGA family [41], manufactured using the Intel 7 process and incorporating second-generation HyperFlex technology, represents a class of advanced SoC FPGAs as shown in Fig. 4 Its "registers-everywhere" paradigm incorporates Hyper-Registers which are bypassable flip-flops distributed along routing paths for fine-tuned timing control, along with Hyper-Retiming and Hyper-Pipelining. Hyper-Retiming repositions these registers to balance path delays and Hyper-Pipelining adds pipeline stages without RTL modifications. This architecture supports clock rates exceeding 1 GHz and enables deep pipelining. The programmable logic fabric contains 34k Adaptive Logic Modules (ALMs), 262 M20K on-chip memories, and 276 variable-precision DSPs arranged in a column-based layout. Dedicated AI Tensor Blocks support INT8, INT16, and mixed-precision operations (e.g., INT4 in early layers and INT8 in later layers) to optimize both efficiency and accuracy. Relative to Xilinx counterparts, the Agilex 3 series demonstrates improved timing closure capabilities and greater pipelining adaptability, while its dual-core Cortex-A55 processors and support for FP32/FP64 operations make it well-suited for executing hybrid AI workloads.

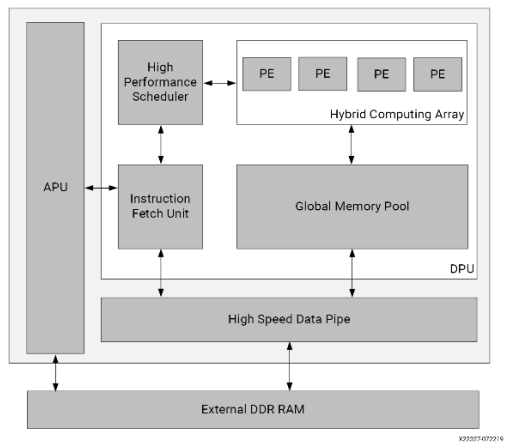

Fig. 3: Top level architecture of DPUCZDX8G [47].

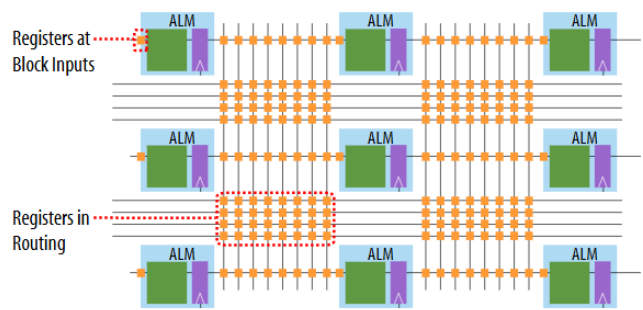

Fig. 4: Hyperflex's "Registers Everywhere" architecture. The orange boxes correspond to the Hyper-Registers in the core fabric [49].

*C. SoC FPGA with hardcore AI engines*

Versal platforms expand the PS and PL architecture by introducing a third compute tier—a 2D mesh of dedicated AI Engine (AIE) and AI Engine-ML tiles. Each tile integrates a SIMD very-long instruction word (VLIW) processor coupled with 128-bit AXI-Stream network-on-chip (NoC) routers [50], [51]. These VLIW cores execute multiple instructions per cycle across scalar, vector, and address-generation pipelines, with vector units supporting numerous data types such as int8/uint8, int16/uint16, cint16, int32/uint32, cint32, BF16, FP16, and int64/uint64. The architecture accommodates vector group widths of 128, 256, 512, or 1024 bits, enabling AI kernels implemented via the AI Engine API to process up to 1024 bits of data each cycle for both integer and floating-point workloads.

In earlier Versal AI Core and Premium series, AIE tiles are equipped with 32 KB SRAM, whereas the AI Edge, HBM, and second-generation AI Core variants incorporate AIE-ML tiles with 64 KB SRAM and integrated DMA engines that enable efficient tensor data transfers. The latest iteration, the AI Engine-ML v2 tiles, used in Versal AI Edge Series Gen 2, further advances this architecture by doubling local SRAM to 64 KB per tile, enhancing the DMA engine to support six simultaneous memory-to-stream and stream-to-memory channels. It upgrades the interconnect fabric with a 64-bit wide AXI4-Stream switch. Additionally, the AIE-ML v2 tiles improve cascade capabilities with horizontal and vertical connections, allowing flexible and scalable array configurations.

The AIE-ML v2 tile architecture comprises several key modules: the AIE-ML v2 core, a memory module with eight 8 KB banks totaling 64 KB, a tile interconnect, and control, debug, and trace units [52]. The AIE-ML v2 core features a scalar datapath, a vector datapath, three address generators, and 16 KB of program memory. The memory module includes a memory interface, DMA functionality, and locks, enabling efficient data movement. The tile interconnect manages incoming and outgoing AXI4-Stream and memory-mapped AXI4 traffic, facilitating seamless communication within the array. The detailed structure of the AIE-ML v2 tile is illustrated in Fig. 5a. Within the AIE-ML v2 array, tiles are arranged in a two-dimensional mesh, with each tile capable of accessing memory modules in all four directions—north, south, east, and west—allowing for flexible and scalable data movement. This architecture enables efficient layer-distributed systolic execution and multi-FPGA scaling, making the Versal platform well-suited for demanding AI workloads. The two-dimensional mesh arrangement of the AIE-ML v2 tiles, showing the inter-tile memory access in all four directions, is depicted in Fig. 5b.

Beyond the structural enhancements, the AIE-ML v2 architecture introduces support for advanced datatypes, notably MX6 and MX9, which offer significant benefits for inference workloads. According to performance data presented at Hot Chips 2024, MX9 achieves up to 99.9% accuracy parity with FP16 on 80% of tested models, while reducing storage by 40% and doubling throughput. MX6 delivers greater than 98% accuracy on more than half of tested models, with 25% lower storage and twice the throughput of dense INT8 [53]. Both formats allow drop-in replacement of quantized weights without retraining, enabling rapid deployment while maintaining model fidelity. Depending on device configuration, AIE-ML v2 delivers between 61 TOPS and 369 TOPS for sparse INT8 workloads, making it a high-performance, power-efficient neural processing unit (NPU) solution for next-generation AI inference at the edge. The AIE MAC, in conjunction with the Versal soft DPU such as the DPUCV2DX8G [54], featuring a systolic array architecture, are tuned for asymmetric per-tensor INT8 weight and activation processing. A compile-time scheduling framework coordinates execution across the AIE tiles and the hard DPU, while a shared-weights buffer minimizes external memory transactions by enabling parameter reuse. Data movement between the AIE, DPU, PL and external DDR is supported by a low-latency SHIM interface and an on-chip NoC fabric, collectively enabling high-throughput, transformer inference tailored for real-time vision workloads, as depicted in Fig. 6.

While the Versal AI Core series does not incorporate high-bandwidth memory (HBM), the Versal HBM variants (e.g., XCVH1582) integrate HBM2e via a 1024-bit interposer connection, offering bandwidths up to 820 GB/s for memory-bound AI applications [55]. Currently, no Intel FPGA family offers an AI Engine subsystem with equivalent architectural capabilities to the Versal AIE framework.

Following the quantitative comparison in Table IV—covering LUT/DSP counts, integrated processors, memory hierarchy, operating frequency, and supported quantization levels—the selection of FPGA architecture for CNN inference depends heavily on target application priorities. LUT-DSP devices provide the most economical option, offering fully flexible quantization (1–16 bits) and deterministic, low-latency data paths, but only moderate throughput in the ≈1–143 TOPS range with low energy efficiency (≈0.2–1 TOPS/W). These are particularly suited to ultra-compact models or vision systems constrained by power budgets. SoC-integrated FPGAs strike a middle ground by combining logic and embedded CPUs, enabling 2–8 TOPS at 1–4 TOPS/W. This makes them effective in domains such as UAV navigation, robotics, and intelligent surveillance where CPU-assisted pre-/post-processing is required. At the high end, AI-centric ACAP platforms such as AMD Versal incorporate dedicated AI Engine arrays, optional HBM2e memory delivering >800 GB/s bandwidth, and mixed-precision compute (INT8/BF16/FP16), enabling >100 TOPS at 8–12 TOPS/W. While their cost is higher, these devices excel in real-time, large-scale deployments where both performance and energy efficiency are mission-critical.

Table V summarizes representative FPGA boards for real-time vision workloads, emphasizing trade-offs between compute density, memory capacity, and price. The AMD ZCU104 offers balanced capabilities with 230 k LUTs and 1,728 DSPs at ≈$1,678. The Versal VC1902 delivers top-tier performance with 400 AI Engines and nearly 2,000 DSPs,

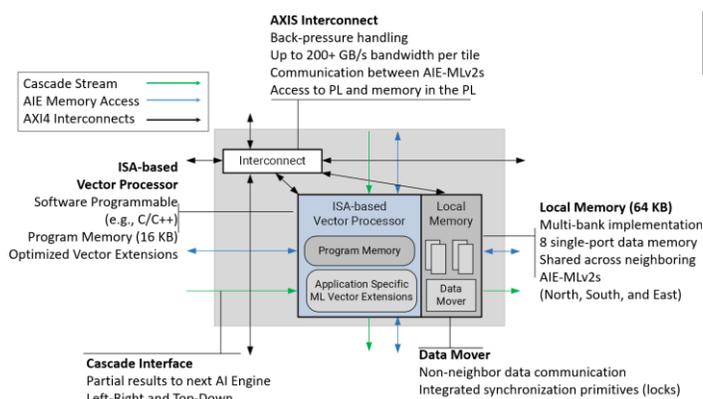
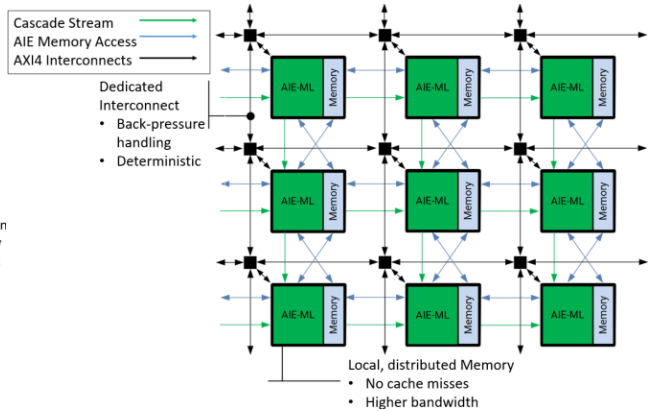

Fig. 5: (a) AI Engine Tile . (b) AI Engine Array from [52]

albeit at ≈$13,000, targeting high-demand AI pipelines. The compact Kria KV260, with 70 k LUTs, suits low-power embedded deployment. Intel's Agilex 7 features high DSP counts and generous on-chip memory, whereas the Arria 10 GX serves as a mid-tier choice for moderate transformer workloads. These boards collectively illustrate the trade-space between performance, integration, and budget.

Recent edge-AI platforms illustrate a clear trade-off between peak compute capability, energy efficiency, and deployment flexibility. The AMD Versal AI Edge Gen 2 [56] delivers up to 185 dense INT8 TOPS (370 sparse TOPS) at ~12–15 TOPS/W, leveraging AI Engines for reconfigurable, sparsity-aware acceleration. NVIDIA's Jetson Thor [57] achieves 2070

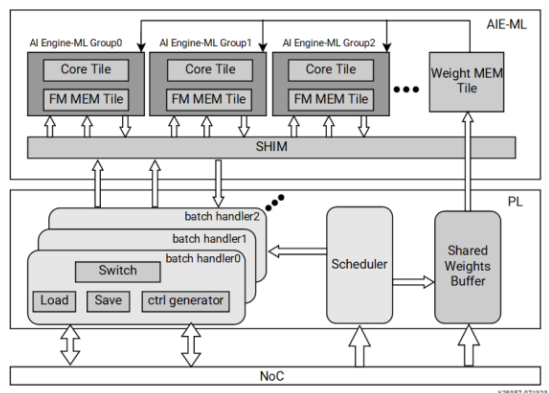

Fig. 6: Block diagram of DPUCV2DX8G [54].

TABLE IV
COMPARISON BETWEEN LUT-BASED, SOC LUT-BASED, AND ACAPS

| FPGA Classifications | LUTs, DSP Slices | Embedded Processors | On-chip SRAM (BRAM / URAM / NoC SRAM) | External DRAM & peak bandwidth | Quantization Support | Typical max clk (fabric / CPU-class core) | INT8 TOPS / TOPS per W |
|---|---|---|---|---|---|---|---|
| LUT-DSP-Based FPGAs | ~10 k – 3 M, 48–12300 DSP48-class slices | None | up to ≈ 455 Mb BRAM + URAM | Soft-IP support for DDR/DDR2/DDR3/DDR4/QDRII+/RLDRAMII/III typical BW < 8 GB/s per channel | Any precision (1–16 bits, custom fixed-point) | 50 – 200 MHz | ~1 – 143 TOPS / ~0.2 – 1 TOPS/W |
| SoC-Based FPGAs | 50K – 2M+ LUTs, ~2 000 – 5 000 DSPs | Hard ARM Cortex-A9/A53 cores (up to quad-core) | 5 – 66 Mb BRAM + ~256 KB OCM | DDR3/4/5, LPDDR4, up to ~1.1 TB/s | INT8, INT16, INT18, INT27 in DSPs (Xilinx), INT4–INT27, FP16, FP32 for Intel devices | 200–400 MHz (PL) / 0.5–1.3 GHz (PS) | ~2 – 8 TOPS / 1 – 4 TOPS/W |
| SoC FPGA with hardcore AI engines | 100K – 2M+ LUTs, Up to ~7K (ACAP total) + AI Engine MACs (up to 400 AIE tiles) | Dual Cortex-A72, dual Cortex-R5, AI Engines (SIMD VLIW cores) | 40 – 245 Mb total on-chip SRAM + AI Engine memory | LPDDR4 / DDR4 (32–64 bit) In-package HBM2e: 32 GB, up to 820–880 GB/s (HBM series) | INT8/INT16/INT32, BF16, FP8, FP16, MX6, MX9, CINT16, CINT32 with AI Engines and DSP58, DSPFP32 and DSPCPLX Modes | 400–600 MHz (PL) / 1.3 GHz (AI Eng.) / 1.5 GHz (A72) | INT8: > 100 TOPS; FP16/BF16: ~ 50 TOPS; TOPS/W ≈ 8–12 |

TOPS at 15.9 TOPS/W, representing the highest raw throughput in this class but with reduced architectural flexibility. Google's Coral M.2 Accelerator with Dual Edge TPU [58] provides 8 TOPS at 2 TOPS/W, targeting ultra-low-power embedded vision deployments. Overall, Versal balances efficiency with adaptability for diverse workloads, Jetson optimizes for maximum throughput in edge-server form factors, and the TPU prioritizes minimal power consumption for always-on, cost-sensitive AI applications.

TABLE V
COMMONLY USED FPGAS FOR REAL-TIME VISION TASKS

| FPGA Board | LUTs / ALMs, FFs | DSP /AI engine | Block RAM (size) | Cost |
|---|---|---|---|---|
| AMD Xilinx Zynq UltraScale+ ZCU104 [59] | 230,400 LUTs, 461,000 FFs | 1,728 | 146 (38Mb) | $1,678 |
| AMD Versal AI Core VC1902 [59] | 899,840 LUTs, 1,799,680 FFs | 1,968 DSPs 400 AI | 34Mb | $13,195 |
| AMD Xilinx Kria KV260 [60] | 70,560 LUTs, 141,120 FFs | 1.2K | 144 | $249 |
| Intel Agilex 7 AGM 032 [61] | 1,100,000 ALMs, 440,0000 FFs | 9,375 | 15,932 (311Mb) | $9,495 |
| Intel Arria 10 GX 1150 [62] | 427,200 ALMs, 1708800 FFs | 1518 | 2,713 (52.96Mb) | $10,019 |

## IV. FPGA-BASED TRANSFORMER AND VLM IMPLEMENTATIONS

### A. Parallelism and FPGA Resource

Transformers and multimodal LLMs demand FPGA mappings that prioritize memory reuse and adaptable parallelism, rather than the strictly local dataflows common to CNNs. Practical strategies often employ a single-load on-chip reuse policy, where weights and activations are fetched only once per block while temporaries remain entirely on-chip. A hybrid tiling method is applied to balance sequence partitioning with head and channel replication, trading BRAM pressure against DSP utilisation. Operator fusion with light approximations is used to execute attention, softmax, and projection in a unified stage. Softmax and LayerNorm are implemented with LUT-friendly approximations to maintain pipeline continuity and eliminate costly off-chip write-backs [63].

As shown in Fig. 7, projection, multi-head self-attention (MSA), and MLP stages in a Transformer encoder can be scheduled contiguously to enable single-load reuse and hybrid tiling. This memory-first scheduling keeps temporaries on-chip, reducing DRAM traffic by an order of magnitude and increasing throughput-per-DSP for ViT inference— particularly on bandwidth-limited edge FPGAs. Operator fusion further preserves pipeline continuity by merging attention, softmax, and output projection into a single stage, with LUT-friendly approximations for softmax and LayerNorm to minimize logic depth and eliminate off-chip write-backs. Combined with systolic or streaming dataflows, these fused operators sustain continuous PE utilization. In designs like the ME-ViT scheduler, linear projection (LP), MHSA, and MLP stages are orchestrated to overlap computation and data movement, minimizing idle cycles.

For multimodal LLMs, FPGA accelerators extend this strategy with cross-attention pipelines that process visual and textual tokens concurrently. Visual patch embeddings are projected and tiled alongside text embeddings, with early-fusion attention layers scheduled in the same PE fabric as standard MHSA. This co-scheduling avoids duplicating hardware for modality-specific blocks and maintains high DSP packing efficiency [64], [65]. By carefully balancing sequence partitioning, head/channel replication, and operator fusion, FPGA designs for Transformers and multimodal LLMs achieve high throughput-per-DSP while staying within tight on-chip memory and bandwidth budgets.

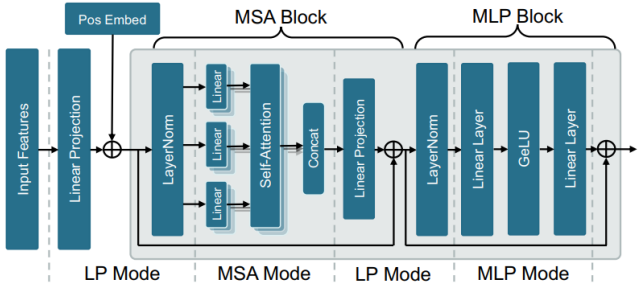

Fig. 7: Block-level scheduling of projection, multi-head self-attention, and MLP stages within a Transformer encoder, demonstrating memory-efficient processing element (ME-PE) mode sequencing for memory-efficient ViT inference on FPGA platforms [63].

### A. Optimization Techniques in FPGAs

Deploying transformers and multimodal LLM workloads on FPGAs requires a different optimization mindset than for CNNs: the implementation must prioritize flexible token/sequence handling, aggressive memory reuse, and careful scheduling of non-local attention computations to stay within tight BRAM, DSP, and DRAM bandwidth envelopes..

#### 1. Spatial Dataflow & Pipelining

Spatial dataflow and careful pipeline design enable FPGAs to sustain high utilization for transformer and multimodal workloads while keeping latency low. The goal is to keep processing elements busy and hide long-latency operations behind useful computation. Recent FPGA accelerators show three practical moves that are especially effective for attention-dominated models:

a) Hybrid-granularity pipelines: Rather than choosing purely coarse (layer-level) or fine (operator-level) pipelines, practical designs combine both. Coarse-stage pipelining yields simple control and low buffer cost for large linear blocks, while fine-grained pipelining inside those blocks removes bubbles and raises clock rates, as shown in Fig. 8a. Architectures that mix these granularities reduce on-chip buffer demand yet maintain steady throughput by inserting only the minimal synchronization points needed between coarse stages. Temporal architectures (often labelled "GeMM" mappings) implement transformers by time-multiplexing a small set of general-purpose PEs to perform the bulk of linear algebra. Each layer is reduced to a sequence of matrix–matrix multiplies and auxiliary ops that are scheduled serially on the same hardware. HG-PIPE demonstrate that hybrid-grained pipelines can substantially reduce pipeline bubbles for ViT-like workloads, as shown in Fig. 8b [66].

b) Decoupled (elastic) stage interfaces and latency hiding: Elastic handshakes or small FIFOs absorb rate mismatches and workload variability (e.g., sequence length, sparsity). Overlapping DMA with compute via ping-pong buffering prefetches the next tile's data while processing the current one. Frameworks like Stream-HLS automate such decoupled, DMA-overlapped pipelines[67].

c) Attention-specific streaming and blocked reductions: Streaming matmuls with blockwise reductions avoid storing full attention matrices. $Q \cdot K^T$ is computed in sub-blocks, partial scores are streamed into an accumulator tree, and softmax/reduction is applied locally. This bounds BRAM use for long sequences and sustains PE throughput, as in tiled/streaming-attention and FlashAttention-inspired FPGA mappings [68].

d) Permutation networks, head shuffles, and on-chip routing: Multi-head attention needs frequent data reordering. Lightweight permutation fabrics (multi-bank scratchpads, crossbar-lite routers, steerable FIFOs) and matched banking strategies (power-of-two banks, staggered accesses) prevent BRAM/URAM hotspotting and maintain backpressure guarantees [69].

Best practices include minimizing initiation interval, managing pipeline depth for timing closure, double-buffering off-chip dependencies, replacing global synchronizations with tree reductions or local rendezvous points [70], [71], [72]. Backpressure mechanisms ensure that if a stage stalls, upstream stages throttle to avoid data overflow or loss [73].

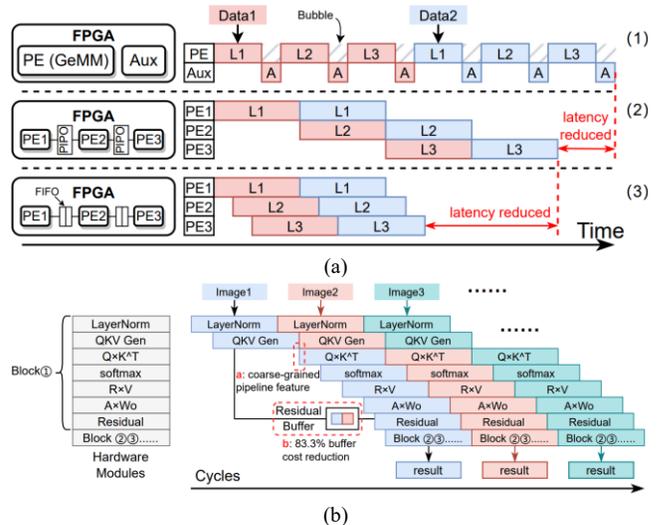

Fig. 8: (a) Difference between pipeline architecture. (1) temporal architecture (2) coarse-grained pipeline architecture (3) fine-grained pipeline architecture. "Aux" is short for auxiliary ops (non-linear functions and off-chip memory access). (b) Hybrid-grained pipeline architecture [66].

#### 2. Precision Scaling & Quantization

Reducing numerical precision is a key lever for shrinking memory footprints and lowering arithmetic and data-movement costs in large Transformer stacks and vision–language models. Well-chosen quantization can preserve task performance while greatly reducing working set size, though the accuracy–efficiency balance is model- and task-specific [74].

There are two strategies for quantization. Post-training quantization (PTQ) quickly converts pretrained models into low-bit formats (e.g., 8-bit or 4-bit) that fit on-chip and ease DDR demands but attention and cross-modal fusion layers can be sensitive to aggressive bitwidth reduction. [75]. To mitigate accuracy loss, hybrid flows apply PTQ broadly, then fine-tune only accuracy-critical layers using quantization-aware training (QAT) or parameter-efficient methods like LoRA [76]. QAT integrates simulated quantization into training to adapt weights to low precision while retaining higher-precision gradients.

Multimodal systems introduce additional challenges due to differing dynamic ranges and outlier patterns between vision and text streams. Dedicated LVLM quantization methods jointly search cross-layer rounding policies to prevent compounding error through fusion; these approaches improve robustness for image–text retrieval and generative tasks relative to naïve layer-wise PTQ [77]. Scaling laws further indicate that optimal bit allocation varies with model size and loss surface curvature, motivating per-chunk or head-aware mixed precision[78], [79].

Hardware-aware quantization translates algorithmic savings into runtime gains by mapping precision schemes directly to FPGA or ACAP architectures. Resource-aware mixed-precision mappings and programmable kernels allow denser MAC unit packing and reduce off-chip bandwidth for Key–Value caches and activation streams[80]. Recent AMD evaluations of the MX6 format show that quantization-aware training (QAT) can recover accuracy lost with naive MX6 application. In the tested flow, weights are first initialized using MX6 quantization, then forward-pass weights and activations are constrained to the MX6 range while emulating the format in FP32 on GPUs. Backpropagation is performed in FP32, and the trained model is finally converted to true MX6 encoding. This staged approach consistently improves accuracy across varied model configurations [53].

In addition, FPGA-oriented approaches like Quasar-ViT use row-wise mixed precision (as in Fig. 9), assigning bitwidths per row (4- or 8-bit) through hardware-aware NAS to balance latency and resources [81]. This enables one-shot training with decomposed low-bit operations, sustaining accuracy and achieving up to 251 FPS on a Xilinx ZCU102 with minimal degradation.

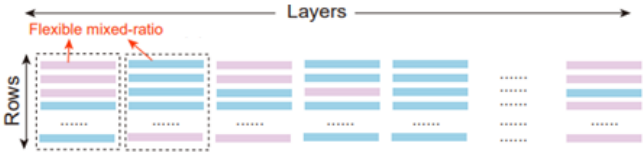

Fig. 9. Illustration of row-wise mixed-precision quantization used in Quasar-ViT. Each row within a weight kernel may be assigned a different bit-width, represented and pink and blue, enabling fine-grained control over accuracy and resource efficiency [81].

*3. DSP Packing and Low-Bitwidth Arithmetic Optimization*

Packing multiple low-precision MAC operations into a single DSP slice is a key optimization for FPGA accelerators of transformer and multimodal LLM inference. Modern DSP primitives, such as Xilinx DSP48E2 and Intel variable-precision DSPs, offer SIMD/packing modes that enable several narrow multiplies per cycle [61], [82]. multiple 8- or 4-bit multiplies within a 27×18 or 54×54 unit can effectively multiply per-DSP throughput without increasing device count [82]. Leveraging these modes requires careful alignment of operand bit-widths, sign handling, and vector widths to DSP input ports so weight and activation formats can be concatenated or interleaved for concurrent evaluation. Quasar-ViT illustrates two schemes on the Xilinx ZCU102 [81] a factor-3 layout placing one 6-bit activation on the 18-bit port and packing three 4-bit weights into the 27-bit port (with small LUT combiners), and a factor-4 layout packing two 6-bit activations with two 4-bit weights. Both require sign extension for upper sub-words and additional LUT logic for packing, trading higher MAC density per DSP for increased routing and timing complexity.

Packed DSP datapaths introduce trade-offs: higher local routing density and I/O fan-out can cause placement congestion and timing closure challenges, while stricter BRAM/URAM organization is needed to align and deliver packed operands within a single cycle. To address this, designers pair DSP packing with tiling and buffer-alignment strategies that ensure co-located vector elements are fetched in one access. Although tools and HLS pragmas can automate some steps, RTL-level control over input packing, carry chains, and pre-adder usage often delivers the best results in high-utilization designs. Quasar-ViT exemplifies this approach, combining row-wise mixed precision with DSP packing to achieve 100–250+ FPS on the ZCU102 within BRAM and DSP limits [81].

For extremely low bitwidths (≤4 bits), two complementary strategies are common. One is aggressive DSP packing, such as mapping multiple INT4 multiplies per DSP, to maximize utilization of existing arithmetic blocks. The other shifts some computation into the LUT fabric, using LUT-based multipliers to surpass the conventional DSP roofline. Recent LUT-multiply designs and micro-scaling formats (e.g., MXInt, MX-style integer formats, and LUTMUL techniques) show that combining LUT arithmetic with DSP packing can increase throughput and area efficiency on midrange FPGAs, albeit with added routing and control complexity [83]. Designers often use a hybrid approach: critical, wide GEMMs remain DSP-mapped with tight packing, while tiny or irregular kernels (or extremely low-bit GEMMs) are implemented in LUT logic to avoid DSP fragmentation and to improve the overall roofline [84].

*4. Pruning and Sparsity*

Pruning refers to the process of removing model parameters or computations that contribute minimally to final accuracy, thereby reducing MAC operations and memory requirements during inference. This removal of parameters induces sparsity, a condition where many elements in the weight or activation tensors are zero, enabling hardware to skip redundant computations. In FPGA-based accelerators, structured sparsity is preferred over unstructured patterns because it preserves regular dataflows that align with systolic arrays and tiled processing elements. Common structured variants include tile-wise pruning, where entire contiguous matrix tiles are removed to allow full PE blocks to be deactivated; head-wise pruning, which eliminates entire attention heads along with their associated Q/K/V projections to reduce attention MAC counts and memory buffers; and channel-wise pruning, which removes full feature or hidden-dimension channels, enabling deallocation or repurposing of compute lanes and BRAM banks.

Structured attention sparsity that limits each token's receptive field is especially effective for long-context workloads: accelerator designs that implement windowed or banded attention achieve large reductions in K/V buffering and DRAM bandwidth while retaining much of the model's expressivity, and these methods translate well into FPGA dataflows because the nonzero pattern is static or semi-static at inference time [85].

Sparsity patterns must be hardware-conscious to avoid irregular memory access. For instance, Parikh *et al.* [86] present an FPGA accelerator for Vision Transformers that combines static block-sparse weight pruning with dynamic token pruning for edge deployment. Fixed-size weight blocks (e.g., 16×16) are pruned via learned score masks for structured sparsity, while runtime pruning retains only top-ranked tokens per head, aggregating dropped tokens into a summary vector. To manage irregular token counts, the design uses a Token Dropping Hardware Module (TDHM) for sorting and re-indexing and a Multi-Level Parallelism Compute Array (MPCA) for structured matrix multiplications, maintaining high PE utilization and data reuse. On DeiT-Small, this approach achieves ~3% accuracy loss, 3.4× computation reduction, 1.6× compression, and 12.8× latency improvement over CPU baselines, highlighting the effectiveness of hardware-aware pruning for real-time FPGA inference.

For deployment, pruning is often coupled with retraining or quantization-aware fine-tuning to recover accuracy. In FPGA systems, this co-optimization ensures that sparsity not only reduces arithmetic load but also aligns with BRAM partitioning and DSP packing, ultimately producing compact, high-throughput designs that can support real-time Transformer and multimodal LLM inference in detection, classification, and tracking pipelines.

## V. Software Toolchains for Real-Time Detection, Classification, and Tracking on FPGAs

Meeting the latency, throughput, and energy constraints of embedded vision applications requires FPGA software environments that integrate model optimization, hardware mapping, and runtime orchestration into a unified workflow. Modern toolchains from AMD/Xilinx and Intel provide such capabilities, enabling rapid deployment of AI workloads ranging from transformer based detectors to large multi-modal language networks in real-time.

### A. Xilinx Toolchain (Vitis Ecosystem)

The AMD/Xilinx Vitis platform offers an end-to-end flow that bridges high-level AI model development with FPGA hardware execution. It integrates High-Level Synthesis (HLS), pre-built DPU overlays, AI framework compatibility, and extensive profiling/debugging utilities into a single development environment. Python APIs and Docker-based deployment options further accelerate prototyping. Key components for vision-based CNN workloads include:

1) **Vitis unified software:** This software combines High-Level Synthesis (HLS) accelerators with C/C++ applications for Zynq UltraScale+ and Kria SOMs and enables hardware/software co-design [87].
2) **Vivado Design Suit:** It provides FPGA synthesis, AXI interconnect management, and IP core integration at the RTL level [88].
3) **Vitis HLS:** It is used to optimize Transformer operators such as matrix multiplications, softmax, and layer normalization via loop unrolling, pipelining, and pragma-based scheduling [89].
4) **Vitis Video Analytics SDK (VVAS):** It provides a GStreamer-based framework to integrate inference, video overlay, and camera capture, with the option to fully offload pipelines to FPGA fabric [90].
5) **Vitis AI:** It is the flagship AI deployment stack for FPGAs [46], as illustrated in Fig. 10, starting from pretrained models in TensorFlow, PyTorch, or Caffe. Models are quantized from FP32 to INT8 using the Vitis AI Quantizer, optionally fine-tuned, and compiled into .xmodel files by the Vitis AI Compiler, which maps operators to DPU resources. The Vitis AI Runtime (VART) exposes C++/Python APIs for execution, while Xilinx Runtime (XRT) manages low-level scheduling and PS–PL data transfers. Pre- and post-processing kernels are available through the Vitis AI Library, and the Vitis AI Profiler provides per-layer latency and throughput breakdowns for iterative tuning. The main limitation is that quantization must start from FP32, restricting certain training flows. Moreover, fixed DPU overlays may limit flexibility for unconventional architectures.

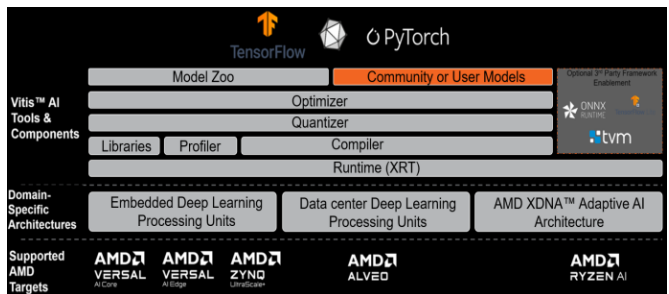

Fig. 10. Vitis AI stack [91].

### B. Intel toolchain

Intel's FPGA AI deployment is driven by the OpenVINO toolkit [92], as shown in Fig. ,which converts trained models from TensorFlow, Keras, or ONNX into an Intermediate Representation (IR) using the Model Optimizer. The Intel FPGA AI Suite DLA Graph Compiler then maps the IR to FPGA logic, generating hardware instructions, weights, and activations according to a user-defined .arch configuration. The flow incorporates compiler-level optimizations such as operator fusion and layer reordering, and supports CPU–FPGA partitioning for hybrid deployments. However, the Intel ecosystem is less transparent in low-level tuning compared to Vitis AI and lacks pre-trained models like model zoo of Vitis AI.

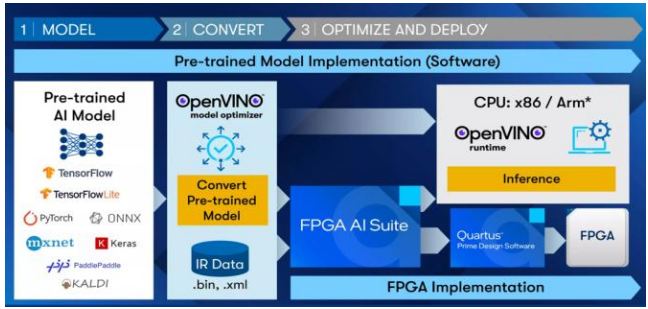

Fig. 11. FPGA AI Inference Development Flow [93].

*C. Other Software Tools*

Specialized open-source toolchains such as FINN [45] and hls4ml [94] target ultra-low-latency and low-precision FPGA implementations. FINN takes quantized neural networks (trained via Brevitas in ONNX format), applies graph-level optimizations such as folding and quantization, and produces deeply pipelined HLS kernels with minimal off-chip memory traffic. hls4ml converts Keras, PyTorch, or ONNX models into synthesizable HLS designs, supporting both Vivado/Vitis and Quartus flows. It provides automated precision analysis, FIFO depth tuning, and latency–resource optimization for real-time workloads.

Recent research has improved FPGA compilation flows without altering proprietary vendor tools. AutoHLS [95], for example, enhances Vitis HLS by using machine learning to guide pragma selection and design-space exploration. By predicting performance and resource usage from sampled kernel configurations, it applies Bayesian optimization to converge on high-performing designs up to 70× faster than manual tuning. Comparable automation tools for Intel FPGA AI flows remain limited.

## VI. TRANSFORMER AND VLM MODELS FOR OBJECT DETECTION, CLASSIFICATION, AND TRACKING AT THE EDGE

*A. Object Detection*

The demand for real-time object detection has intensified with the growth of safety-critical and interactive systems. While convolutional detectors remain dominant in embedded hardware deployments, transformers and VLMs are increasingly attractive due to their superior representation learning and open-vocabulary capabilities. The challenge lies in reconciling their high computational density and memory requirements with the tight throughput, power, and resource limits of FPGAs.

Early designs like DETR [19] replaced region proposal mechanisms with direct set-based predictions, removing complex post-processing such as NMS, but dense global attention and large key–value caches led to high latency, memory use, and irregular data movement, limiting FPGA suitability. Later variants, including Deformable DETR [96] and RT-DETR [97], mitigate quadratic attention complexity via multi-scale or sparse attention: Deformable DETR attends to a small set of points around reference positions, reducing computation and key–value storage, while RT-DETR streamlines the encoder and employs hybrid attention to achieve over 100 FPS on GPUs, enhancing FPGA compatibility. Hierarchical transformers such as Swin Transformer further improve spatial granularity and reduce bandwidth by restricting self-attention to local windows with periodic shifts, enabling efficient attention kernel tiling and pipelining on DSP and AI Engine arrays for FPGA deployment.

The most recent advancement, RF-DETR (Region-Focused DETR) [98], extends the Deformable DETR backbone by introducing region-aware query initialization and Token-to-Region Attention (TRA), as shown in Fig. 12. Using a Region Proposal Encoder (RPE), it generates region tokens embedding spatial and semantic information, and TRA restricts cross-attention to each query's region, reducing redundant key–value lookups and K/V cache sizes. Single-scale processing simplifies on-chip buffers and lowers off-chip memory transactions, producing predictable memory access patterns suitable for FPGA. These features make RF-DETR more hardware-friendly while maintaining competitive accuracy, with strong potential for efficient FPGA deployment when combined with pruning, quantization, and parallelization strategies.

Despite the architectural advances of transformers, no fully integrated end-to-end real-time transformer detector has yet been demonstrated on FPGA in peer-reviewed literature. The key obstacles remain the scaling behavior of multi-head attention, large intermediate activations, and decoder-side post-processing such as bipartite matching. Current FPGA research has focused on accelerating encoder blocks or attention submodules while offloading detection heads to CPU or GPU co-processors. Achieving a complete real-time pipeline for transformer-based detection on a single FPGA will require co-design of attention algorithms, KVQ memory systems, and dataflow scheduling to fit within the tight on-chip resource and bandwidth constraints of these devices.

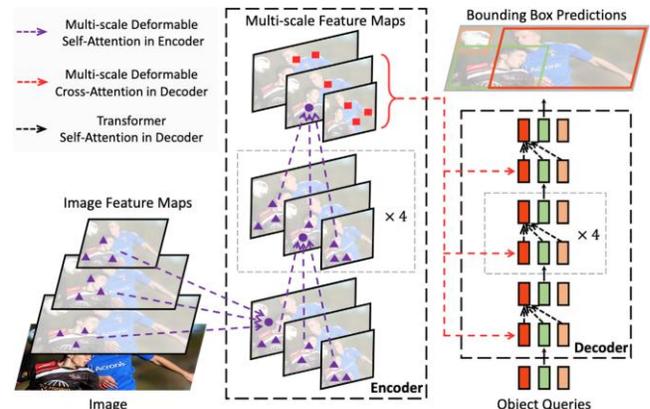

Fig. 12. Architecture of Deformable DETR (RF-DETR is built on this architecture) [98]

LVLMs can leverage textual context to enrich visual understanding and detection. For instance, Zang et al. [99] present ContextDET, that employs a visual encoder, an LLM decoder for language context, and a visual decoder to produce bounding boxes based on text, enabling "generate-then-detect" object identification. The recent models, such as Grounding DINO, and quantization-optimized systems like Q-VLM, extend transformer frameworks to handle multi-modal inputs, integrating visual tokens with textual queries for open-vocabulary recognition. Vision encoders in these systems can be implemented on FPGA using similar optimizations as pure transformer detectors, including blocked attention, operator fusion, and per-layer bit-width scaling. Cross-attention between vision and language features, however, increases the sequence length and, consequently, the memory footprint, making it more difficult to achieve fully on-chip execution without HBM support. Large-scale language reasoning components, common in advanced VLMs, exceed the capacity of most embedded FPGAs even after aggressive quantization, leading to split-execution designs where the FPGA accelerates the vision path and a host processor handles the language model.

Another recent advancement in this space is Paligemma-2 [100], introduced by Google in late 2024, which combines a lightweight Gemma-2 LLM backbone with a SigLIP vision encoder in an encoder–decoder configuration, as shown in Fig. 13. The SigLIP encoder divides the input image into patches, converts them into visual tokens, and projects them into the Gemma-2 text embedding space through a learned adapter, enabling joint processing with textual tokens. The Gemma-2 decoder alternates multi-head self-attention and cross-attention layers, enabling bi-directional fusion between visual and textual streams for open-vocabulary detection, caption-conditioned classification, and multimodal reasoning. From an FPGA perspective, the SigLIP encoder is relatively amenable to acceleration using Vision Transformer techniques such as blocked attention, patch embedding tiling, and mixed-precision quantization. However, the Gemma-2 decoder remains computationally demanding; cross-attention inflates key–value cache sizes and stresses on-chip memory bandwidth, especially for devices without HBM. As such, a practical future deployment strategy is to map the SigLIP encoder and projection layers onto the FPGA while offloading Gemma-2 decoding and language reasoning to a host processor, with full on-chip execution feasible only on AI-engine–rich FPGAs with aggressive pruning and sub-4-bit quantization.

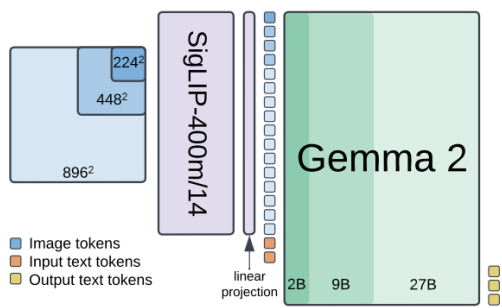

Fig.13: Architecture of Paligemma-2, combining a SigLIP-400m/14 visual encoder with the Gemma-2 language model. The encoder processes images of 224px², 448px², or 896px² resolution into 256–4096 patch tokens, which are linearly projected and concatenated with input text tokens. The fused sequence is autoregressively processed by Gemma-2 for multimodal reasoning and output generation [100].

*B. Object Classification*

Real-time image classification focuses solely on assigning a single label or a small set of labels to an input image, in contrast to object detection, which must also localize objects via bounding boxes or segmentation masks. This lack of spatial localization requirements makes classification pipelines inherently less computationally intensive than detection. However, modern transformer-based classifiers still face significant performance challenges on FPGAs due to the computational complexity discussed in section II. Recent FPGA accelerators for classification showcase diverse yet complementary strategies:

1) **Single-load, pipelined PEs:** ME-ViT [63] introduces a memory-efficient Vision Transformer acceleration. By employing PEs that buffer weights on-chip and reuse intermediate activations, ME-ViT eliminates repeated off-chip transfers as disussed in section IV. A Multiple ME-PE architecture (Multi-PE) is also developed by arranging parallel ME-PE along with a scheduler to coordinate data traffic between them.

2) **Hybrid-grained pipelined architecture**: HG-PIPE [66], as discussed earlier, achieves exceptional throughput ($\approx$ 7118 images/s—~2.8× faster than a V100 GPU) and significantly improved resource efficiency.

3) **Mixed-precision execution:** TATAA [101] focuses on adaptive arithmetic for Transformer operators, dynamically switching between low-precision integer formats for compute-intensive linear layers and higher-precision arithmetic for sensitive non-linear functions such as softmax, LayerNorm, and GELU. This mixed-precision execution reduces hardware resource usage while maintaining classification accuracy. By tailoring precision to each operation's sensitivity, TATAA achieves substantial power savings without degrading inference throughput.

4) **Mixture-of-Experts (MoE) accelerators:** UbiMoE [102] optimizes Mixture-of-Experts Vision Transformers via streaming attention kernels and reusable linear blocks, tuned across FPGA platforms using a heuristic search. It achieves up to 3.35× throughput and ~1.5× higher energy efficiency over prior FPGA designs. CoQMoE [103] extends this by integrating a dual-stage quantization pipeline—balancing complex and hardware-friendly quantizers—for MoE models. It delivers ~5.35× throughput and over 80 % energy reduction versus SOTA MoE accelerators, with <0.3 % classification accuracy loss.

5) **Ultra-Low-Latency and Automated FPGA Deployment Frameworks:** EQ-ViT [104] uses a spatially heterogeneous accelerator on AMD Versal ACAP, pairing AIE vector cores for matrix multiplications with PL logic for nonlinear and element-wise functions, as shown in Fig. 14. Multiple specialized accelerators run concurrently with on-chip data forwarding and fine-grained pipelining to maximize utilization. INT8-friendly Softmax (INT-Softmax$2^n$) and GeLU (I-GeLUImp) replace costly floating-point operations with LUT and bit-shift logic, reducing DSP/LUT usage. A co-optimized mapping framework balances resource allocation, workload partitioning, and quantization, achieving 0.56 ms ViT inference—over 13× faster than FPGA baselines—while improving accuracy via activation-aware QAT. Additionally, the CAT framework [105] automates transformer accelerator generation for AMD FPGAs by mapping compute-intensive matrix multiplications to AIE cores and control-heavy or element-wise functions to PL. Using parameterized hardware templates, on-chip buffering, and streaming interconnects, CAT minimizes off-chip memory access and sustains continuous dataflow. A design space exploration process tunes parallelism, tiling, and quantization for the target device. When compared to existing ViT deployments on AMD Alveo U50, ZCU102, and VCK190 boards, CAT-based accelerators achieve throughput gains of 97.9×, 42.6×, and 1.37×, respectively. These improvements are matched by energy efficiency increases of 62.2×, 5.85×, and 1.37×. Additionally, for the BERT model, migrating from a baseline accelerator on the Zynq Z-7100 to a CAT-based implementation on the VCK5000 results in a 169.2× throughput boost and a 50.09× rise in energy efficiency.

Wu et al. compared TAATA accelerator performance and power efficiency of transformers and LLMs on FPGAs and GPU platforms. As shown in Figure 15, TATAA delivers higher normalized power efficiency on FPGA than on GPUs, especially on LLMs, proving that when these extended to VLMs, will provide better efficiency than on GPUs.

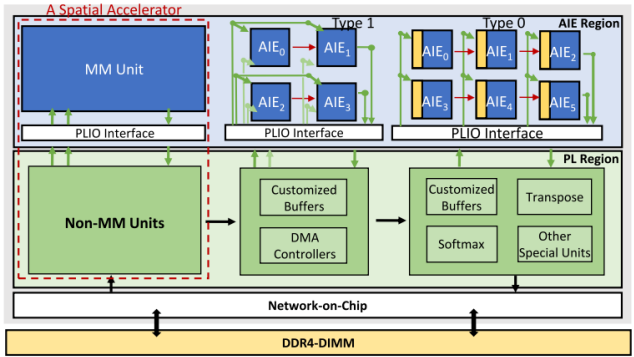

Fig.14: Architecture of EQ-ViT [104].

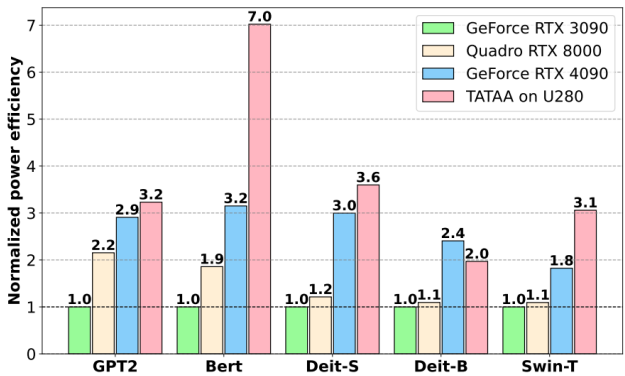

Fig.15: Power efficiency comparison of TATAA on U280 FPGA and on other GPU [101].

Table VI compares different transformer models for classification on FPGAs. Deit-tiny on HG-PIPE accelerator and ViT on CAT framework provides lowest latency with a trade-off on power. The latter has highest throughput while $M^3$ViT-T on CoQMoE accelerator offers low resources. Although most literatures uses Alveo U280/U200 accelerator cards that contains UltraScale+ FPGAs which belongs to LuT-DSP based FPGAs, Versal FPGAs offer top throughput and latency which makes them suitable for real-time classification applications.

Recent advances in VLMs for classification on GPUs have demonstrated significant improvements in multimodal understanding by integrating large-scale transformer architectures. Models such as Flamingo 2 and BLIP-3 employ cross-modal attention mechanisms that fuse visual embeddings with textual context, enabling open-vocabulary and prompt-driven classification with high accuracy. Architecturally, these models rely on deep multi-layer transformer stacks with extensive use of self-attention and cross-attention modules, processing sequences of visual tokens alongside language tokens. While highly effective on GPU platforms using their massive parallelism, high memory bandwidth, and flexible scheduling, these architectural features present substantial challenges for FPGA deployment. The dynamic sequence lengths, irregular memory access patterns from cross-attention, and large intermediate activations strain FPGA on-chip resources and complicate efficient pipelining and parallelization. Furthermore, the frequent model updates and adaptations for new tasks require runtime flexibility that current FPGA design flows do not readily support.

No peer-reviewed literature reports a fully integrated real-time classification system based on VLMs implemented entirely on FPGA hardware. Beyond the aforementioned computational complexity and resource demands, the scarcity of dedicated hardware-software co-design frameworks tailored to VLM classification, the limited availability of FPGA-optimized multimodal transformer kernels, and the lack of comprehensive benchmark datasets for FPGA-targeted VLM evaluation further hinder research progress in this area. Addressing these gaps will require new methodologies that unify model simplification, dynamic dataflow management, and adaptive hardware reconfiguration to harness FPGA capabilities effectively for real-time multimodal classification.

C. Real-Time Tracking

Real-time object tracking involves continuously identifying and monitoring objects within video sequences with very low delay, a critical capability for applications such as autonomous driving, robotics, security monitoring, and augmented reality [115], [116]. In complex or densely populated environments, effective tracking facilitates early detection of unusual activities in surveillance systems and supports proactive navigation decisions for vehicles [116]. Moreover, tracking plays a vital role in vehicle navigation by forecasting the future locations of moving objects within the surroundings.

Transformers have recently gained traction in the domain of real-time object tracking, with models like HFFTrack and DeforT pushing the boundaries of accuracy by incorporating hybrid frequency features and deformable attention modules. These architectures leverage deep layers of self-attention and cross-attention to effectively model spatial and temporal relationships across video frames. On GPUs, these models benefit from high throughput and flexible memory hierarchies, enabling smooth real-time performance. However, the direct translation of these transformer-based trackers to FPGA platforms remains largely unexplored. The reasons are multifaceted: multi-head attention demands substantial on-chip memory and compute resources, while the variability in sequence lengths and dynamic data access patterns complicate the design of efficient, low-latency dataflows on FPGAs.

TABLE VI
COMPARISON BETWEEN RECENT REAL-TIME CLASSIFICATION METHODS ON FPGA

| Ref. | Model | FPGA | Freq. (MHz) | Frame Rate (FPS) | Accuracy | Image Size | Latency (ms) | Through-put (GOPS) | Power (W) | BRAM | LUTs/ ALMs | DSP | FFs |
|---|---|---|---|---|---|---|---|---|---|---|---|---|---|
| [63] | Multi-PE ME-ViT | Alveo U200 | 300 | 132.04 | - | 224×224 | 37.86 | 2682 | 31.8 | 576 | 384,000 | 2048 | 264,000 |
| [101] | Swin-T on TAATA | Alveo U280 | 225 | - | 79.98 | - | - | 2512.3 | 10.8 | 1472 | 724.9 | 4352 | 1154.9 |
| [66] | Deit-tiny on HG-PIPE | Versal VCK190 | 425 | 7118 | 71.05 | 256×192 | 0.136 | 17795 | 46.7 | 1006.5 | 669k | 312 | - |
| [102] | ViT-S on UbiMoE | Alveo U280 | 250 | - | - | - | 11.66 | 789.72 | 31.36 | 974 | 316.1k | 3413 | 385.9k |
| [103] | M3ViT-T on CoQMoE | ZCU102 | 300 | - | 84.89 | - | 6.47 | 386.3 | 9.83 | 383 | 156.1k | 1754 | 198.2k |
| [103] | M3ViT-S on CoQMoE | Alveo U280 | 250 | - | 84.89 | - | 9.16 | 1004.3 | 33.7 | 696.5 | 311.6k | 2635 | 454.1k |
| [105] | ViT on CAT | Versal VC190 | 300 for PL and 1.33GHz for AIE | - | - | - | 0.129 | 30,279.75 | 61.464 | 706 | 261.4K | - | 262.4K |

Moreover, the fast pace of model evolution and the need for adaptable inference engines challenge the relatively static nature of FPGA hardware.

Currently, no published work documents a fully integrated real-time transformer tracker implemented on FPGA. Beyond architectural hurdles, this void is compounded by a scarcity of dedicated hardware-software co-design tools optimized for tracking transformers, a lack of FPGA-optimized transformer primitives, and the absence of comprehensive datasets tailored for evaluating FPGA-based tracking solutions. To bridge this gap, future research must focus on creating streamlined transformer variants amenable to FPGA constraints, innovating adaptive hardware scheduling techniques, and developing robust frameworks that enable efficient end-to-end real-time tracking on FPGA devices.

Recent advancements in VLMs have extended their capabilities to real-time object tracking, particularly through models like TrackVLA [106]. Fig. 16 compactly depicts TrackVLA's end-to-end VLA pipeline: a vision encoder extracts patch features that are grid-pooled into fine tokens for the current frame and coarse tokens for recent history, then a small projector maps these visual tokens into the LLM latent space and concatenates them with the language prompt and a special [Track] token. The combined token sequence is forwarded through a shared LLM backbone (a Vicuna-class model), whose conditional embedding is routed to one of two heads—an autoregressive language head for recognition/VQA or an anchor-based diffusion action head for tracking. The diffusion head denoises clustered trajectory anchors into scored waypoint sequences, from which the top trajectory is decoded into low-level control commands. Deploying such transformer-based VLMs for real-time tracking on FPGA platforms presents significant challenges. The extensive use of multi-head self-attention mechanisms and large-scale language models leads to substantial memory and computational demands, which are difficult to accommodate within the limited resources of typical FPGA devices. Additionally, the dynamic nature of language-conditioned tracking requires flexible and adaptive processing capabilities that current FPGA architectures do not readily support.

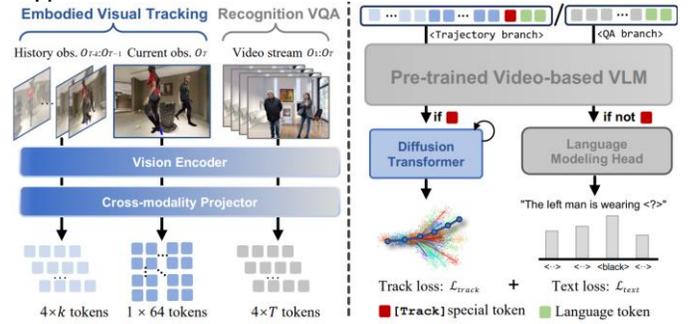

Fig.16: Pipeline of TrackVLA [106].

Consequently, there is a notable absence of peer-reviewed literature addressing the real-time deployment of transformer-based VLMs for object tracking on FPGA platforms. This gap is attributed to several factors: the complexity and size of transformer models, which exceed the memory and computational capacities of most FPGAs; the lack of hardware-optimized components for multimodal transformers; and the absence of standardized benchmarks for evaluating FPGA-based VLM tracking systems. Overcoming these challenges necessitates the development of novel FPGA architectures that can efficiently handle the demands of multimodal processing and the creation of comprehensive evaluation frameworks tailored to FPGA implementations.

## VII. DESIGN TRADE-OFFS AND CHALLENGES IN FPGA-BASED VISION INFERENCE

This section summarizes practical design trade-offs and deployment challenges that are especially relevant when mapping vision transformers and multimodal language–vision models to FPGA fabrics.

Selecting the target hardware platform fundamentally shapes the available design choices. Lower-cost LUT–DSP-based FPGAs provide flexibility for custom datapaths and low latency but lack the memory capacity to store large activations or key–value caches. Mid-range SoC devices integrate CPU fabrics with programmable logic, easing orchestration and enabling moderate throughput. At the high end, ACAP or HBM-equipped boards offer the bandwidth and specialized compute engines required for low-latency, multimodal workloads, though they introduce higher cost and greater thermal and design complexity.

A primary challenge in transformer-based vision models is the rapid growth of memory demand as token counts increase. The central bottleneck lies in maintaining key–value matrices and intermediate activations close to the compute units. Addressing this can involve provisioning large on-package memory, designing streaming or fragmented attention schemes to limit the active working set, or modifying the attention mechanism itself through sparsity or blocking. Each choice shifts the balance between latency, accuracy, and engineering complexity.

Keeping the compute fabric fully utilized requires careful coordination of data movement and processing. Strategies such as tiled streaming, operator fusion, and the use of elastic FIFOs can minimize redundant transfers and improve MAC utilization. However, these gains come at the cost of intricate buffer management, complex permutation logic, and more demanding control schemes, all of which increase design effort and verification time.

Performance can be further improved by adopting low-bit quantization and packing multiple narrow operations into a single DSP unit, significantly boosting throughput density. To maintain accuracy under these constraints, model-aware quantization and often retraining are necessary. Exploiting structured sparsity aligned to hardware resources allows efficient skipping of computations, whereas irregular sparsity patterns typically fail to deliver practical benefits due to packing inefficiencies.

The development process itself involves trade-offs between rapid prototyping and maximum efficiency. HLS tools and vendor-provided runtimes can shorten development cycles but may obscure fine-grained optimizations essential for high-utilization designs. In contrast, manually optimized RTL or finely tuned HLS yields better resource efficiency but requires longer and more complex development and verification cycles, making engineering effort an important part of the performance equation.

Designing FPGA-based accelerators for vision–language models adds further layers of complexity. The dual-modality nature means that both visual and textual processing pipelines must be optimized and synchronized, often requiring different data representations, precision formats, and throughput characteristics. The vision encoder typically involves convolutional or transformer-based feature extraction that benefits from high parallelism and streaming dataflows, while the language decoder demands efficient handling of sequence dependencies and dynamic memory access patterns. Balancing these heterogeneous compute and memory requirements on a single fabric is a key architectural challenge.

VLM inference often needs to handle variable-length text prompts and dynamically generated output sequences, which adds runtime control complexity. While fixed-shape batching simplifies scheduling, real-world applications such as image captioning or visual question answering produce highly variable workloads. Supporting this flexibility in FPGA designs typically requires elastic buffering, adaptive scheduling, and fine-grained synchronization between processing stages, all of which consume extra logic and control resources.

From a deployment standpoint, many VLM use cases require interaction with external processing units for tasks such as pre/post-processing, database access, or API communication. This multi-device orchestration increases the importance of low-latency interconnects and efficient data marshaling between heterogeneous components. Additionally, given the rapid pace of model evolution, FPGA designs must be modular enough to accommodate new embedding formats, attention variants, and architectural updates without requiring a complete hardware redesign.

Evaluation and benchmarking for FPGA-based VLM accelerators also lag their vision-only counterparts. Multi-modal workloads demand coordinated metrics for latency, throughput, accuracy, and even qualitative assessment of generated outputs. The absence of standardized, FPGA-oriented evaluation frameworks for VLM inference makes fair comparison difficult, complicating architectural decisions and slowing the refinement process.

Future advances in FPGA-based vision–language and transformer acceleration are likely to be driven by tighter algorithm–hardware co-design. Emerging attention mechanisms such as linearized, block-sparse, or state-space–inspired architectures can significantly reduce memory footprints and sequence-processing complexity, enabling higher utilization of limited on-chip resources.

On the deployment side, we can expect a shift toward modular, reconfigurable overlays that abstract low-level hardware details while supporting fast adaptation to evolving model architectures. Hardware-aware compilers and automated design space exploration tools will streamline mapping of diverse VLM and transformer workloads, enabling rapid iteration without sacrificing efficiency. Moreover, deployment on Intel FPGAs needs to be explored along with improved Intel software. Standardized FPGA-centric benchmarking for multimodal inference will improve comparability and guide optimization efforts. Also, scheduling tasks between FPGA, GPU and NPU may results in better inference. Together, these developments point toward accelerators that are not only faster and more efficient, but also resilient to the rapid evolution of model architectures, ensuring sustained relevance in real-world, latency-critical applications.

## VIII. Conclusion

FPGA-based acceleration of transformers and vision–language models offers a unique balance of flexibility, energy efficiency, and low-latency processing, making it an attractive choice for edge and embedded AI deployments. However, the diverse computational patterns, large memory footprints, and evolving architectures of these models present significant design and deployment challenges, from hardware resource allocation and dataflow optimization to runtime flexibility and multi-modal integration. Addressing these constraints requires not only careful device selection and microarchitectural tuning but also a co-design philosophy that aligns algorithmic innovation with hardware capabilities.

As model architectures continue to advance, future FPGA solutions must prioritize scalability, portability, and adaptability while sustaining high utilization and predictable performance in real-world scenarios. The integration of algorithm-aware compression, structured sparsity, modular overlays, and standardized evaluation frameworks will be key to bridging the gap between research prototypes and robust, deployable systems. With these advancements, FPGAs can remain a competitive and sustainable platform for next-generation multimodal AI workloads.